\documentclass[article]{aa}
\pdfoutput=1

\usepackage[english]{babel}
\usepackage{multicol}
\usepackage{amsmath}
\usepackage{amsfonts}
\usepackage{dsfont}
\usepackage{amsxtra}
\usepackage{hyperref}
\usepackage{amssymb}
\usepackage{upgreek}
\usepackage{changes}

\usepackage{multirow}
\usepackage{url}
\usepackage{float}

\usepackage{graphicx,epsfig}
\usepackage{dcolumn}

\onecolumn

\usepackage{xspace} 

\usepackage{color}
\usepackage{xcolor}

\usepackage[switch]{lineno}
\usepackage{lipsum}

\newcommand{\be}{\begin{equation}}
\newcommand{\ee}{\end{equation}}
\newcommand{\bea}{\begin{eqnarray}}
\newcommand{\eea}{\end{eqnarray}}
\newcommand{\beaa}{\begin{eqnarray*}}
\newcommand{\eeaa}{\end{eqnarray*}}
\newcommand{\ba}{\begin{array}}
\newcommand{\ea}{\end{array}}
\newcommand{\bi}{\begin{itemize}}
\newcommand{\ei}{\end{itemize}}
\newcommand{\ben}{\begin{enumerate}}
\newcommand{\een}{\end{enumerate}}

\newcommand{\lb}{\label}

\newcommand{\adeg}{^\circ\!\!\,}

\definecolor{darkgreen}{rgb}{0.0, 0.7, 0.0}

\begin{document} 

   \title{NGC\,3894: a young radio galaxy seen by \textit{Fermi}-LAT}

   \author{G. Principe \thanks{\email{giacomo.principe@inaf.it}}
          \inst{1}
          \and
          G. Migliori 
          \inst{1,2}
          \and
          T. J. Johnson
          \inst{3}
          \and
          F. D'Ammando 
          \inst{1}
          \and
                  M. Giroletti 
          \inst{1}
          \and
          M. Orienti 
          \inst{1}
          \and
          C.~Stanghellini 
          \inst{1}
          \and   
          G.~B.~Taylor
          \inst{4}
          \and
          E.~Torresi 
          \inst{5}
          \and
          C. C. Cheung 
          \inst{6}          
          }

   \institute{
               {INAF - Istituto di Radioastronomia, Bologna, Italy}
            \and
               {Dip. di Fisica e Astronomia, Universit\`{a} di Bologna, Bologna, Italy}
            \and
            {College of Science, George Mason University; Resident at Naval Research Laboratory, Washington, DC, USA}
            \and
              {University of New Mexico, Albuquerque, NM, USA}
            \and
            {INAF - Osservatorio di Astrofisica e Scienza dello Spazio di Bologna, Bologna, Italy}
            \and
               {Naval Research Laboratory, Space Science Division, Washington, DC, USA}
             }

   \date{Received November 5, 2019; accepted February 28, 2020}

\abstract{According to radiative models, radio galaxies may produce $\gamma$-ray emission from the first stages of their evolution. However, very few such galaxies have been detected by the \textit{Fermi} Large Area Telescope (LAT) so far.}
{NGC\,3894 is a nearby (z =0.0108) object that belongs to the class of compact symmetric objects (CSOs, i.e., the most compact and youngest radio galaxies), which is associated with a $\gamma$-ray counterpart in the Fourth \textit{Fermi}-LAT source catalog.
Here we present a study of the source in the $\gamma$-ray and radio bands aimed at investigating its high-energy emission and assess its young nature.}
{We analyzed 10.8 years of \textit{Fermi}-LAT data between 100 MeV and 300 GeV and determined the spectral and variability characteristics of the source. Multi-epoch very long baseline array (VLBA) observations between 5 and 15 GHz over a period of 35 years were used to study the radio morphology of  NGC\,3894 and its evolution.
}
{NGC 3894 is detected in $\gamma$-rays with a significance $>9 \sigma$ over the full period, and no significant variability has been observed in the $\gamma$-ray flux on a yearly time-scale. The spectrum is modeled with a flat power law ($\Gamma=2.0\pm0.1$) and a flux on the order of $2.2 \times 10^{-9}$ ph cm$-2$ s$^{-1}$.
 For the first time, the VLBA data allow us to constrain with high precision the apparent velocity of the jet and counter-jet side to be $\beta_\mathrm{app,NW}=0.132\pm0.004$ and $\beta_\mathrm{app,SE}=0.065\pm0.003$, respectively.}
{\textit{Fermi}-LAT and VLBA results favor the youth scenario for the inner structure of this object, with an estimated dynamical age of $59\pm5$ years. The estimated range of viewing angle ($10\adeg<\theta<21\adeg$) does not exclude a possible jet-like origin of the $\gamma$-ray emission.
}

\keywords{Galaxies: evolution - galaxies: nuclei - galaxies: individual (NGC 3894, 1146+596, UGC 06779) - galaxies: jets - radio continuum: galaxies - gamma-rays: galaxies}

\maketitle

\section{Introduction}
\lb{sec:intro}
The extragalactic $\gamma$-ray sky is dominated by blazars \citep{2015A&ARv..24....2M}, for which the $\gamma$-ray emission is favored by the small jet inclination angle and beaming effect. However, the increasing amount of data collected by the Large Area Telescope (LAT) on board the \textit{Fermi Gamma-ray Space Telescope} \citep{2009ApJ...697.1071A} allows us to also investigate other classes of objects in the $\gamma$-ray sky \citep{2018A&A...614A...6S}. A small percentage, $\sim2\%$ in the fourth catalog of $\gamma$-ray AGN \citep[4LAC,][]{2019arXiv190510771T}, is represented by radio galaxies (or misaligned AGN), which have larger jet inclination angles ($> 10 ^{\circ}$) and a smaller Doppler factor ($\delta \leq2-3$). 

Compact symmetric objects (CSOs, i.e., the most compact and youngest radio galaxies), with their symmetric and subkiloparsec radio structure, are important objects because they are expected to be the progenitors of the extended radio galaxies \citep{1996ApJ...460..612R}.
In principle, CSOs with a powerful jet could produce $\gamma$-ray emission up to the GeV band through inverse Compton scattering of the ultraviolet (UV) photons from the accretion disk by the electrons in the compact radio lobes \citep[][see the second reference for a discussion of hadronic models]{2008ApJ...680..911S,2011MNRAS.412L..20K}. 
However, systematic searches of CSOs in $\gamma$-rays have so far been rather unsuccessful \citep{2016AN....337...59D}. Dedicated studies have reported a handful of detections of CSO candidates \citep{2011ApJ...738..148M, 2014A&A...562A...4M}. Of these, the nearby (z=0.0144) radio galaxy NGC\,6328 (also called PKS\,1718-649) is the most probable case \citep{2016ApJ...821L..31M}, which has recently been confirmed in the Fourth \textit{Fermi}-LAT Source Catalog  \citep[4FGL,][]{2019arXiv190210045T}.

The 4FGL catalog reports the association of a newly detected $\gamma$-ray source (4FGL J1149.0$+$5924) with the radio galaxy NGC\,3894 (also known as UGC\,06779 or 1146$+$596), which is also classified as a CSO \citep{2000ApJ...534...90P}.
NGC\,3894 \citep[$z = 0.010751$,][]{1991rc3..book.....D}\footnote{At this luminosity distance (50.1 Mpc), 1 mas corresponds to 0.23 pc.}  is one of the least radio-luminous CSOs; it is hosted in an elliptical galaxy whose optical continuum emission is dominated by starlight \citep{2001AJ....122..536P}. On the basis of the [OIII]/H$\beta$ versus [NII]/H$\alpha$ and [OI]/H$\alpha$, \citet{2004A&A...413...97G} classified the source as a low-ionization emission-line region (LINER), possibly indicating a low-power central engine.
Observations with very long baseline interferometry (VLBI) over 15 years allowed the identification of the radio core and provided evidence for twin parsec-scale jets \citep{1998ApJ...502L..23P}. Both the jets were estimated to be mildly relativistic ($v \sim 0.3 c$) and oriented away from the line of sight \citep[$\theta \sim 50\adeg$,][]{Taylor1998}.

In this paper, we investigate the $\gamma$-ray and radio properties of NGC\,3894. We analyze 10.8 years of \textit{Fermi}-LAT data between 100\,MeV and 300\,GeV as well as several epochs of VLBA radio data, extending the \cite{Taylor1998} coverage from 15 years to $\sim$35 years, in order to investigate the high-energy characteristics of NGC\,3894 and verify its young nature.

Throughout this article, we assume $H_{0} = 70$ km s$^{-1}$ Mpc$^{-1}$ , $\Omega_{M} = 0.3$, and $\Omega_{\Lambda}=0.7$ in a flat Universe.

\section{\textit{Fermi} data and analysis}
\lb{sec:data}
The LAT is a $\gamma$-ray telescope that detects photons by conversion into electron-positron pairs. It has an operational energy range of 20\,MeV to 300\,GeV and beyond. The LAT is comprised of a high-resolution converter tracker (for direction measurement of the incident $\gamma$-rays), a CsI(Tl) crystal calorimeter (for energy measurement), and an anticoincidence detector to identify the background of charged particles \citep{2009ApJ...697.1071A}.

We performed a dedicated analysis of the \textit{Fermi}-LAT data with the goals of (1) confirming the detection and spatial association of 4FGL\,1149.0+5924 with the CSO NGC\,3894, (2) determining its $\gamma$-ray spectral properties, and (3) investigating its temporal behavior. We collected $\sim$10.8 years of Pass 8 LAT data \citep{2013arXiv1303.3514A} that were collected between August 4, 2008, and June 20, 2019, which means that we consider a more extended dataset than in the 4FGL (8 years).
We selected events that were reprocessed with the P8R3\_Source\_V2 instrument response functions (IRFs) \citep{2018arXiv181011394B} in the energy range between 100\,MeV and 300\,GeV because below 100 MeV the source is not detected \citep{2018A&A...618A..22P}. The region of interest (ROI) has a radius of about 20$^{\circ}$ that is centered on the position of the $\gamma$-ray source 4FGL\,J1149.0+5924, as reported in the 4FGL catalog \citep[$\rm R.A.,decl. (J2000) = 177\fdg25,59\fdg42$,][]{2019arXiv190210045T}.

The binned likelihood analysis (which consists of model optimization, localization, and a study of the spectrum and variability) was performed with Fermipy\footnote{http://fermipy.readthedocs.io/en/latest/} \citep{2017arXiv170709551W}, a python package that facilitates analysis of data from the LAT with the \textit{Fermi} science tools, of which the version 11-07-00 was used. The maps were created with a pixel size of $0.1\adeg$. 

Gamma rays with a zenith angle larger than 90$\adeg$, as well as photons below 300 MeV from the point spread function (PSF) 0 event type\footnote{A measure of the quality of the direction reconstruction is used to assign events to four quartiles. 
Gamma rays in Pass 8 data can be separated into four PSF event types: 0, 1, 2, and 3, where PSF0 has the largest PSF and PSF3 has the best.}, were excluded in order to eliminate most of the contamination from secondary $\gamma$-rays from the Earth’s limb \citep{2009PhRvD..80l2004A}. 
The model used to describe the sky includes all point-like and extended LAT sources that are located at a distance $<25^{\circ}$ from the source position, as listed in the 4FGL, as well as the Galactic diffuse and isotropic emission.
For these two latter contributions, we made use of the same templates\footnote{https://fermi.gsfc.nasa.gov/ssc/data/access/lat/ \\BackgroundModels.html} as were adopted to compile the 4FGL.
For the analysis we first optimized the model for the ROI (fermipy.optimize), then we searched for the possible presence of new sources (fermipy.find\_sources), and finally, we relocalized the source (fermipy.localize).
We investigated the possible presence of additional faint sources that are not in 4FGL by generating test statistic (TS) maps. The test statistic is the logarithmic ratio of the likelihood of a model with the source being at a given position in a grid to the likelihood of the model without the source, TS=2log($\frac{likelihood_{src}}{likelihood_{null}}$) \citep{1996ApJ...461..396M}. We found three new sources that we added to our model. The best-fit positions of these new sources are R.A., decl. (J2000) = (182.62$\adeg$, 59.15$\adeg$), (180.96$\adeg$, 58.31$\adeg$) and (177.36$\adeg$, 62.77$\adeg$), with a 95\% confidence-level uncertainty R$_{95}$ = 0.07$\adeg$.

We left free to vary the diffuse backgrounds and the spectral parameters of the sources within 5$^{\circ}$ of our target.
For the sources in a radius between 5$^{\circ}$ and 10$^{\circ}$ only the normalization was fit, while we fixed the parameters of all the sources within the ROI at larger angular distances from our target.
The spectral fit was performed over the
energy range from 100 MeV to 300 GeV.

\section{\textit{Fermi}-LAT results on NGC 3894}
\label{fermi_results}
The results of the localization analysis present a significant excess at a TS = 98, corresponding to a significance $>9\sigma$, centered on the position (R.A.,decl.(J2000)) = ($177.25\adeg\pm 0.02\adeg$,$59.41\adeg\pm 0.02\adeg $), 95\% confidence-level uncertainty R$_{95}=0.05\adeg$. This is compatible with the position of 4FGL\,J1149.0+5924.

\begin{figure}[h]
\centering
\includegraphics[width=10cm]{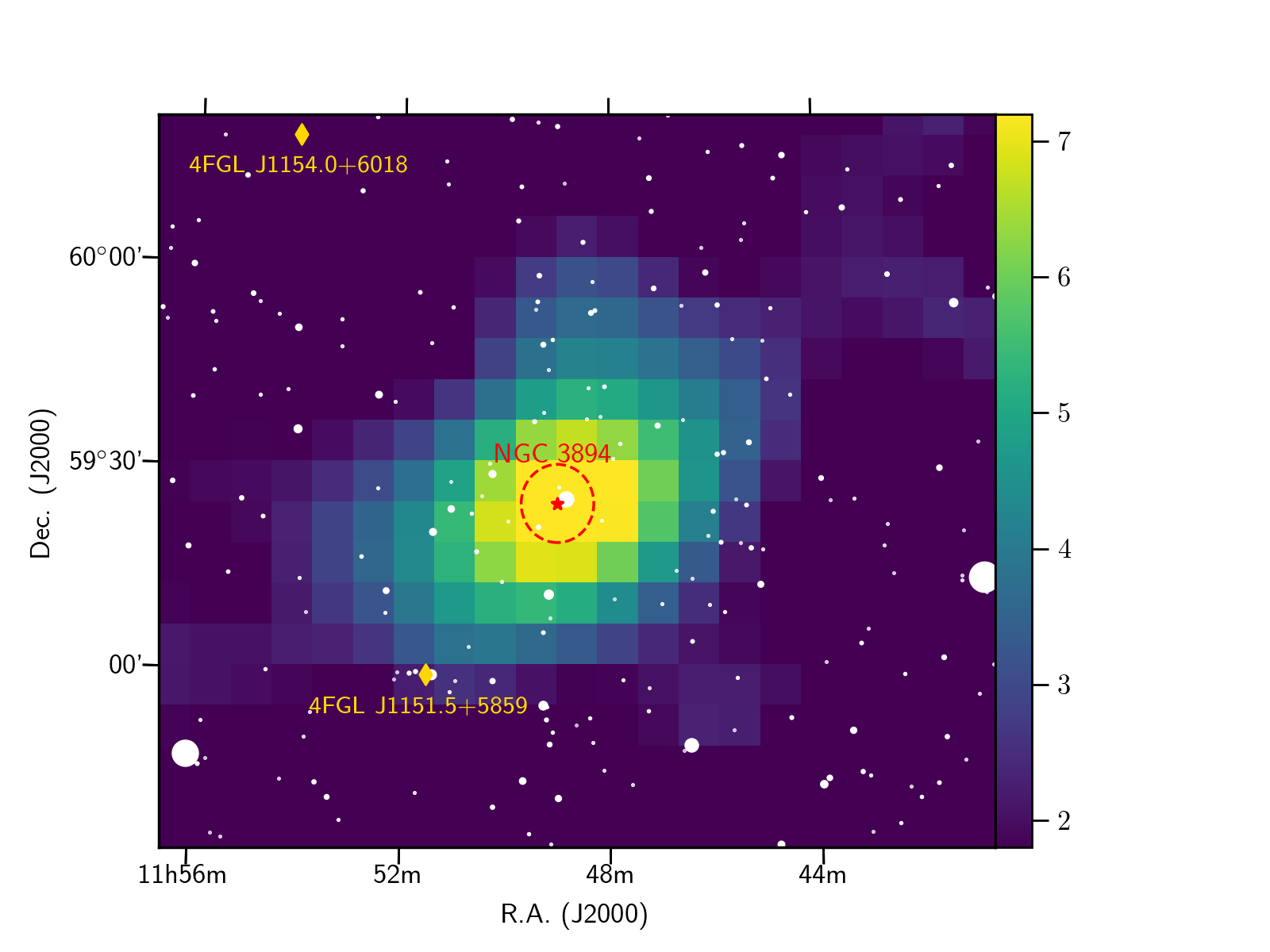}
\caption{\small \label{fig_excess_map} \textit{Fermi}-LAT TS map (in sigma units) above 3 GeV, the red star and dashed circle represent the central position and the  95\% confidence-level uncertainty R$_{95}=0.05\adeg$ of the $\gamma$-ray source, respectively. White dots show radio sources from the NVSS survey scaled depending on their flux.}
\end{figure}

NGC\,3894, with an offset of $0.024\adeg$ and a flux density of 482 mJy, is the nearest bright radio source (in the NVSS survey) to 4FGL\,J1149.0+5924, making it the best candidate radio counterpart.
No other NVSS sources are detected within the $R_{LAT,95}$ down to 4 mJy.
NGC\,3894 has also been proposed as the most likely counterpart in the 4LAC using both the Bayesian method (Probability$\sim$0.998) based on spatial coincidence and the likelihood ratio method (Probability$\sim$0.968) based on radio flux density.
Fig. \ref{fig_excess_map} shows the \textit{Fermi}-LAT TS map (in sigma units) above 3 GeV for the region around NGC\,3894.

\subsection{\textit{Fermi}-LAT spectral energy distribution}
\begin{figure}[h]
\centering
\includegraphics[width=8cm]{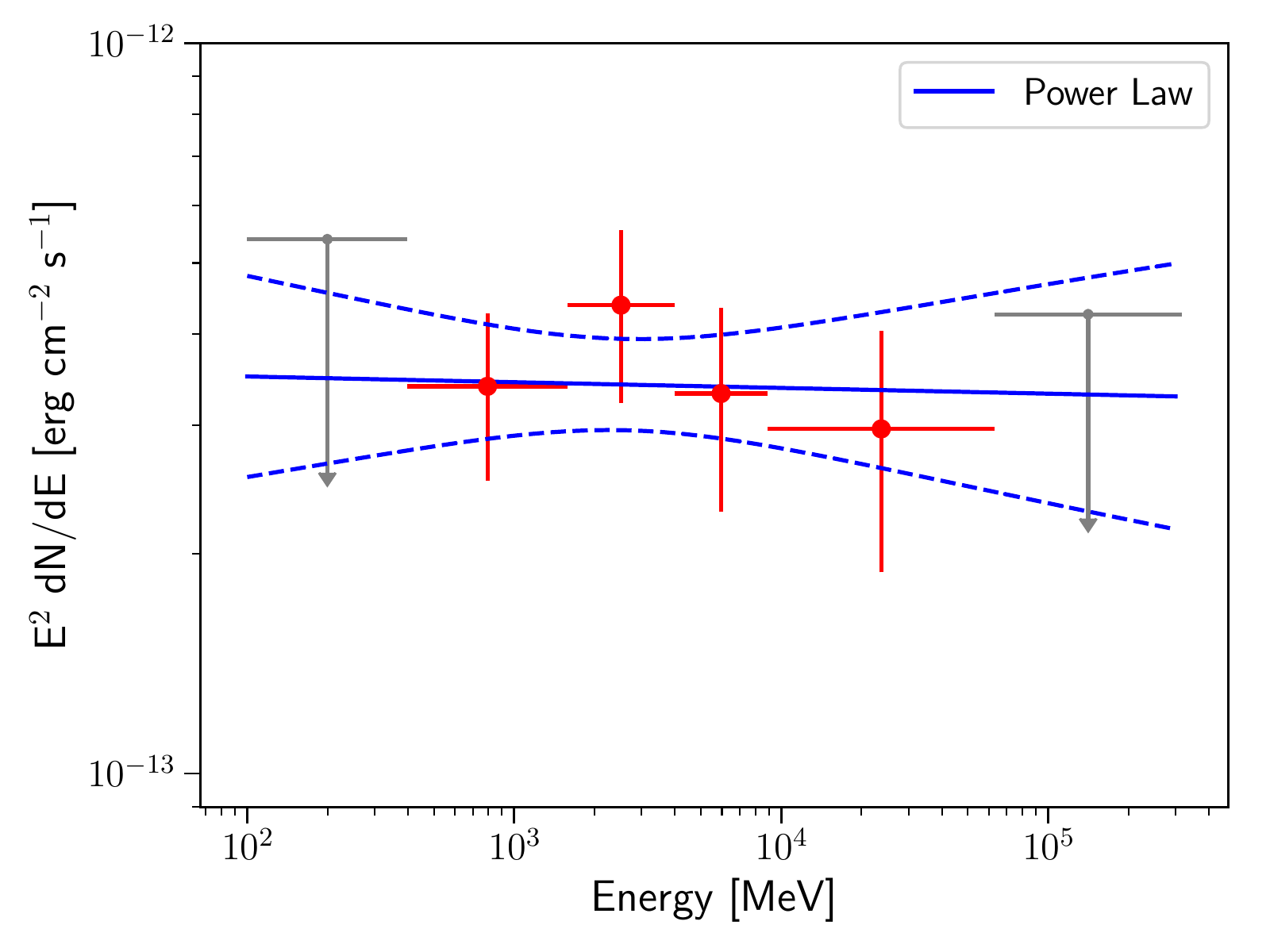}
\caption{\small \label{sed_fermi} \textit{Fermi}-LAT spectrum of NGC\,3894 obtained using 10.8 years of data between 100\,MeV and 300\,GeV. The spectrum was fit with a power law (blue line). The 1 $\sigma$ upper limit is reported when TS < 10.}
\end{figure}

We modeled the spectrum of the source with a power-law function ($\dfrac{dN}{dE} = N_{0} \times (\frac{E}{E_{b}})^{- \Gamma}$). 
The best-fit results for the $\gamma$-ray source associated with NGC\,3894 are $\Gamma = 2.01 \pm 0.10$ and $N_{0} =( 2.14 \pm  0.39) \times 10^{-13}$ (MeV cm$^{-2}$ s$^{-1}$) for $E_0=1$GeV, which are in agreement with the 4FGL results for this source ($\Gamma_\mathrm{4FGL} = 2.06 \pm 0.12$). 
Figure \ref{sed_fermi} shows the spectral energy distribution (SED) we obtained here using 10.8 years of \textit{Fermi}-LAT. 

\subsection{\textit{Fermi}-LAT study of emission variability}
\label{fermi_lightcurve}
\begin{figure}[h]
\centering
\includegraphics[width=\columnwidth]{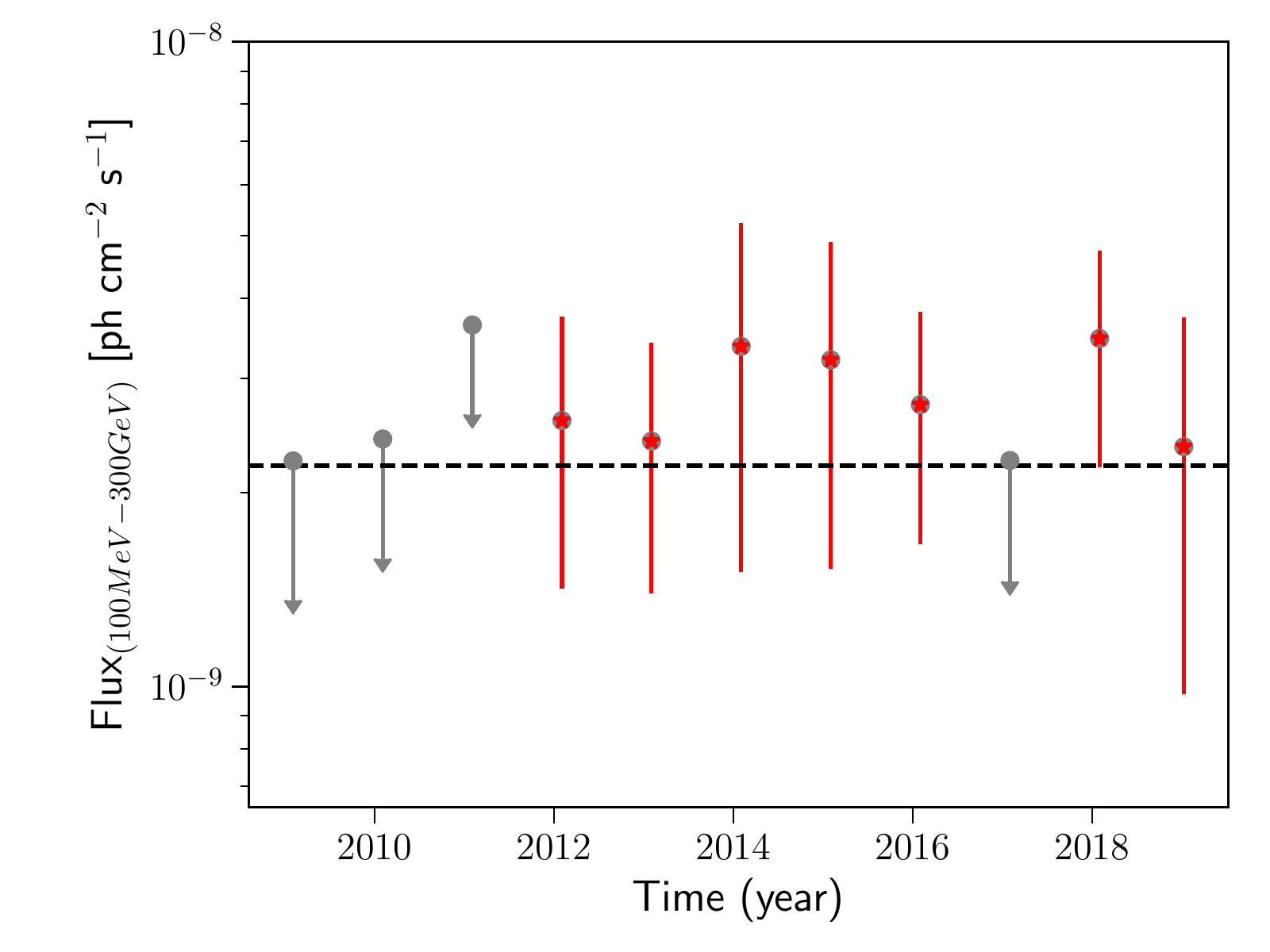}
\caption{\small \label{fig_lightcurve} \textit{Fermi}-LAT one-year binned light curve of NGC 3894. The dashed line represents the averaged flux for the entire period $Flux_{10.8\,years}=2.2\times 10^{-9}$ ph cm$^{-2}$ s$^{-1}$. }
\end{figure}

Over a period of eight years,   the source presents a variability index of $TS_{\mathrm{var}}$=0.97 \footnote{A measure of the variability index.} in the 4FGL, where a value of $TS_{\mathrm{var}}>$ 18.5 is used to identify a variable sources at 99\% confidence level.
In our analysis we extended the study of the $\gamma$-ray emission variability to a period of 10.8 years in order to investigate possible presence of high activity in the last three years. 
To perform a study of the $\gamma$-ray emission variability of NGC\,3894, we divided the \textit{Fermi}-LAT data into time intervals of one year; ten months was used for the last bin. For the light-curve analysis we fixed the photon index to the value obtained for 10.8 years of data, $\Gamma = 2.01$ (see Section \ref{fermi_results}), and left only the normalization free to vary. The 95\% upper limit was reported in each time interval with TS $<10$. 
Figure \ref{fig_lightcurve} shows the light curve for time bins of about one year, using 10.8 years of \textit{Fermi}-LAT data. 

\noindent 
No indication of variability is observed in the $\gamma$-ray flux of NGC\,3894: a test of the uniform distribution for the averaged flux $F=(2.2\pm 1.0) \times 10^{-9}$ ph cm$^{-2}$ s$^{-1}$ returns a value of $\chi^{2}/\mathrm{ndf}$=2.3/10.
Our results extend for a longer period what has been obtained in the 4FGL and confirm that there is no indication of $\gamma$-ray flux variability of NGC\,3894, as seen for most misaligned sources.

\section{New radio results on NGC 3894}
To complement the datasets presented by \citet{Taylor1998}, we downloaded VLBA archival data for five epochs of observations spanning 1995 to 2017 between 5 and 8 GHz. Table \ref{table_vlbaobs} presents the VLBA archival datasets we used. 

\begin{table}
        \centering
                \begin{tabular}{c c c l}
                        \hline  \hline
                          Project& date & $\nu$ (GHz) & Notes \\
                        \hline
                        BR026& 22 Mar 1995 & 8.4 & NRAO \\
                        BV019& 24 Aug 1996 & 5 & NRAO \\
                        BT094& 30 Jun 2007 & 8.4 & NRAO \\
                        RV122& 25 Apr 2017 & 8.4 & Astrogeo \\ 
                        RV122& 04 Aug 2017 & 8.6 & Astrogeo \\ 
                        \hline
                \end{tabular}
        \caption{VLBA archive data. Col.~(4) lists NRAO data downloaded from the archive and calibrated as described in the text, and Astrogeo are calibrated visibility data downloaded from \url{astrogeo.org} and that were only analyzed with Modelfit in Difmap. The RV122 experiment is described in \citet{Petrov2009}. }
                \label{table_vlbaobs}
\end{table} 

 We carried out amplitude and phase calibration in a standard way following the guidelines given in Appendix C of the Astronomical Image Processing System (AIPS) cook book \citep{2003ASSL..285..109G}. Data were imaged in AIPS after some iterations of phase self-calibration.  
Model fitting of the calibrated $(u,v)$-data\footnote{The Fourier transform of the sky brightness distribution observed by interferometers.} was performed with the task MODELFIT available in the software package DIFMAP \citep{1994BAAS...26..987S}.

\begin{figure*}
    \centering
        \includegraphics[width=12cm]{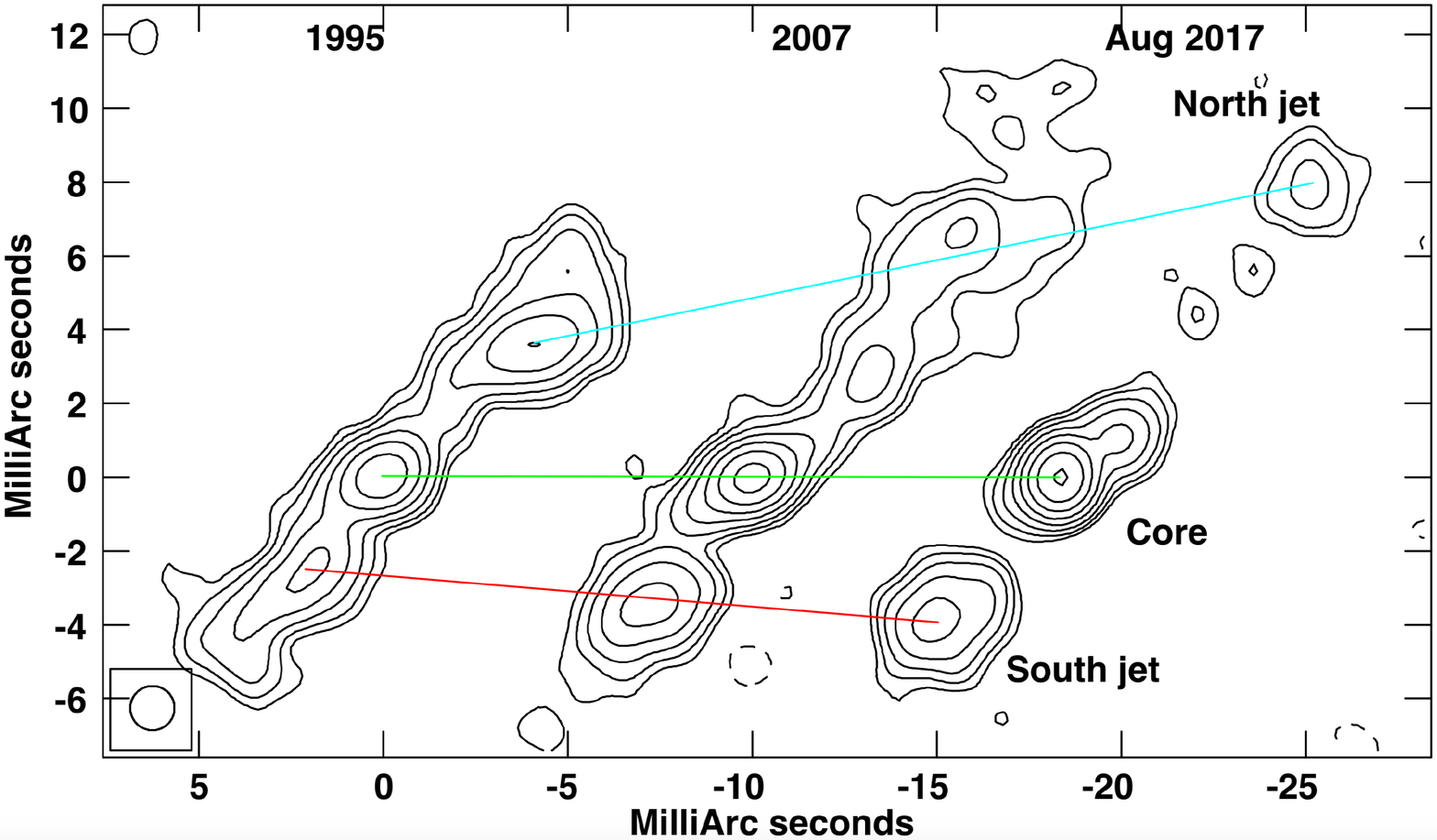}
        \caption{Comparison of three VLBA epochs at 8.4 GHz in 1995, 2007, and August 2017 (left to right). Images are spaced horizontally by a distance proportional to the time separation. Contours are traced at $\pm(1,2,3,8,16,\dots)\times 1.5$ mJy beam$^{-1}$. The colored lines trace the central position of the NW jet (blue), SE jet (red), and core (green), respectively. All the images are convolved with the same circular beam (shown in the bottom left corner), whose full width at half-maximum is 1.2 mas. }
        \label{fig_radioimages}
\end{figure*}

The radio source appears two-sided with respect to a central brightest flat-spectrum component identified as the radio core \citep[see Fig.~\ref{fig_radioimages}; for a spectral index map, see also Fig.~2-I in][]{Tremblay2016}. The structure is resolved into different components that are elongated over $\sim20$ mas at position angle $\sim-40^\circ$ (measured positive north-to-east), corresponding to an overall projected size of $\sim$4 pc. Some more distant and diffuse emission is recovered, particularly on the southern side, in the 5 GHz datasets, in agreement with what has been reported at 1.4-1.6 GHz by \citet{1998ApJ...502L..23P} and \citet{Taylor1998}.
The data have rather different distributions of the visibilities, therefore a precise association and comparison of the distinct components seen at different epochs or frequencies is challenging. In our model-fit procedure, we used a minimal number of discrete Gaussian components.

\subsection{Constraints on geometry from fits to radio images}
\begin{figure}
    \centering
    \includegraphics[width=9cm]{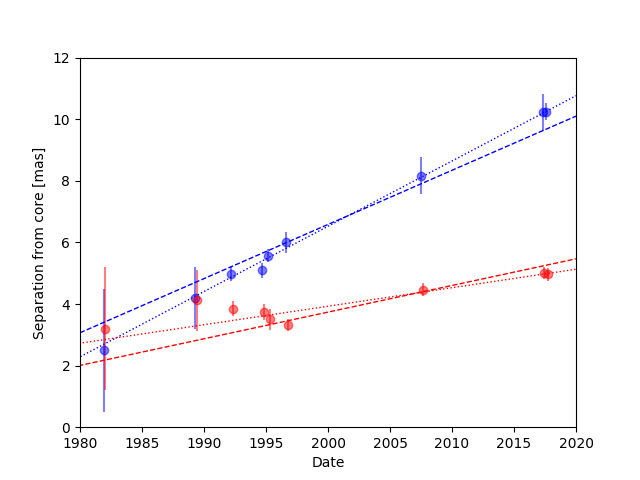}
    \caption{Separation from core for the NW (in blue) and SE (in red) components. For each set of points (see also Table \ref{table_modelfit}), the dotted line shows an unconstrained linear fit to the data, while the dashed line includes the additional condition that the ejection dates of the two components are consistent with each other.}
    \label{fig_separation}
\end{figure}

The outermost NW and SE external components are consistently detected across epochs and frequencies. 
In Table \ref{table_modelfit} and Fig.~\ref{fig_separation} we show the separation of the NW and SE component from the central core as a function of time. We assume that the core is stationary across epochs and that any opacity-related shift in its position is negligible in comparison with the observational uncertainty on the position of each component. We only considered very close frequencies (5 and 8.4 GHz) and rather extended components, therefore this is a sensible assumption. 

 \begin{table*}
                \centering
                \begin{tabular}{c c c c c c c} 
                \hline\hline                       
                         Epoch & $\nu$ & $r_\mathrm{NW}$ & $\theta_\mathrm{NW}$ & $r_\mathrm{SE}$ & $\theta_\mathrm{SE}$ & Ref. \\
                             & (GHz) & (mas) & ($^\circ$) & (mas) & ($^\circ$) &\\
                        \hline  
                        1981.9 & 5   &  $2.5\pm2.0$ & -49.0 & $3.2\pm2.0$ & -225.0 & 1 \\
                        1989.3 & 5   &  $4.2\pm1.0$ & -48.4 & $4.1\pm1.0$ & -221.3 & 2 \\
                        1992.2 & 5   &  $5.0\pm0.3$ & -42.8 & $3.9\pm0.3$ & -223.1 & 2 \\
                        1994.7 & 5   &  $5.1\pm0.3$ & -46.8 & $3.7\pm0.3$ & -219.1 & 2 \\
                        1995.2 & 8.4 &  $5.6\pm0.2$ & -47.1 & $3.5\pm0.3$ & -218.0 & 3 \\
                        1996.6 & 5   &  $6.0\pm0.3$ & -46.9 & $3.3\pm0.2$ & -217.1 & 3 \\
                        2007.5 & 8.4   &  $8.2\pm0.6$ & -41.7 & $4.5\pm0.2$ & -219.0 & 3 \\
                        2017.3 & 8.4 & $10.2\pm0.6$ & -42.0 & $5.0\pm0.2$ & -221.4 & 3 \\
                        2017.6 & 8.6 & $10.2\pm0.3$ & -40.4 & $5.0\pm0.2$ & -220.1 & 3 \\
                        \hline
                \end{tabular}
                \caption{Separation of NW and SE components with respect to the core. Col.~(7) lists the references. 1: \citet{Wrobel1985}, 2: \citet{Taylor1998}, and 3: this work. Uncertainties are derived from the component size along the direction to the core, divided by a factor $5$, except for the first two epochs: for 1981.9, in which the confusion is severe and the $(u,v)$-coverage sparse, we adopt an uncertainty of 2 mas, while for 1989.3, which has better quality and somewhat larger separation, we adopt 1 mas.}
                \label{table_modelfit}
\end{table*}

As was noted by \citet{Taylor1998}, the source is not well described by simple Gaussian components, and caution should be used in interpreting the results, particularly for the SE side component. However, based on our now $\sim35$-year long time baseline, we can determine the motion on the NW side with high precision and use it as a prior for a self-consistent scenario describing the kinematics of NGC~3894. In particular, we successfully fit our data with a model in which the two components were both ejected at $t_\mathrm{ej}=1961\pm5$, corresponding to a dynamical age of $59 \pm 5$ years, and they are separating from each other at a rate of $(0.269\pm0.007)\ \mathrm{mas} \ \mathrm{yr}^{-1}$, or in units of the speed of light, $\beta_\mathrm{app}=v_\mathrm{app}/c=0.202\pm0.005$. In this scenario, the individual speeds of the two components are $\beta_\mathrm{app,NW}=0.132\pm0.004$ and $\beta_\mathrm{app,SE}=0.065\pm0.003$.

We can use these results and the well-known relation for the arm-length ratio

\begin{equation}
    D=\frac{\mu_\mathrm{NW}}{\mu_\mathrm{SE}}=\frac{d_\mathrm{NW}}{d_\mathrm{SE}}= \frac{1+\beta\cos\theta}{1-\beta\cos\theta} \, ,
\end{equation}

\noindent where $\mu$ (mas/years) is the proper motion of each component (jet and counter jet) with respect to the core, to set a first constraint on $\beta\cos\theta\ge0.34\pm0.03$, because our measurements yield $D=2.03\pm0.12$.

The other complementary constraint that is usually provided by the brightness ratio is more  ambiguous in our case because the flux density of each component varies from epoch to epoch, and their ratio does not have a constant value or even a well-defined trend.   It is not clear how systematic effects due to the changes in $(u,v)$-coverage can be separated from more physical causes, which involve not only the values of $\beta$ and $\theta,$ but also the spectral index, the fluid and pattern velocity difference, the cases of continuous jet versus discrete component, the relevance of adiabatic losses in the observer's frame, and others.

\begin{figure}
    \centering
    \includegraphics[width=9cm]{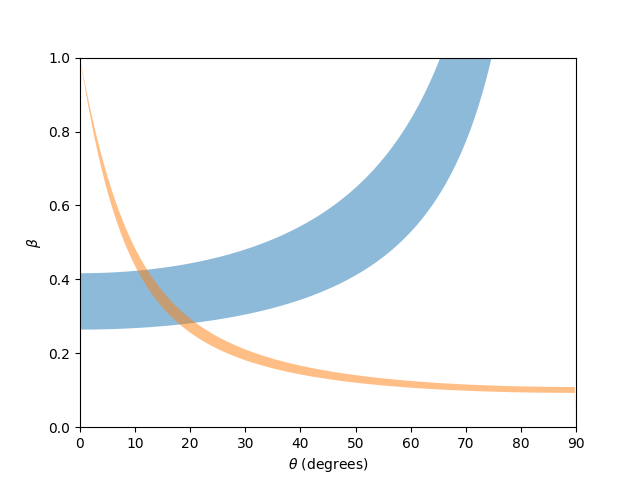}
    \caption{$(\theta,\beta)$-plane, with constraints from the jet and counter-jet arm-length ratio (in blue) and the separation velocity (in orange); the colored regions indicate the $3\sigma$ confidence ranges.}
    \label{fig_beta_theta}
\end{figure}

We can instead use the overall separation velocity 

\begin{equation}
    v_\mathrm{sep}=\mu_\mathrm{sep}D_\theta(1+z)=\frac{2\beta c \sin{\theta}}{1-\beta^2\cos^2\theta} \, ,\end{equation}

\noindent where $D_\theta$ is the angular size distance and $\mu_\mathrm{sep}$ is the summed proper motion of jet and counter jet with respect to the core, to obtain an independent relation between $\theta$ and $\beta$. In Fig.~\ref{fig_beta_theta} we show the regions of the $(\theta,\beta)$ parameter space that are allowed by our measurements. These results indicate a misaligned nature, with a viewing angle in the range $10^\circ \le \theta \le 21^\circ$, and the jet velocity in the range ($0.28\le\beta\le0.43$ or $1.04\le \textrm{Lorentz factor} \le1.11$). The corresponding values of the Doppler factor decrease from $\delta\sim1.6$ in the top left part of the selected area to $\delta\sim1.3$ in the bottom right.

The proposed scenario is not the only viable one. We could relax the assumption of  simultaneous ejection of the NW and SE components or invoke different interactions on the two sides as the cause of the observed asymmetries.  While these scenarios would not allow us to set any meaningful constraints and are therefore less interesting to explore, the unambiguous motion with subluminal velocity in the NW jet and the presence of emission on both sides of the core argue strongly for an overall misaligned and only mildly relativistic jet.

\section{Discussion and conclusion}
Our analysis confirms the detection (TS>90) of faint $\gamma$-ray emission ($L_{\gamma}\sim 6 \times 10^{41}$ erg s$^{-1}$) from the radio galaxy NGC\,3894.
The radio VLBA analysis based on the extended time baseline yields a somewhat smaller viewing angle ($10\adeg<\theta<21\adeg$) than the previous estimates of \citet{Taylor1998}, nevertheless, it is still suggestive of a more misaligned nature in comparison with typical $\gamma$-ray blazars, an overall compact size, and low or moderate bulk motion.
The misaligned scenario for NGC\,3894 is further supported by a number of multi-wavelength properties reported in the literature. The main results supporting this scenario are the lack of polarization in the radio images \citep{Tremblay2016}, the absence of flaring activity in both radio and $\gamma$-ray data \citep[only slow variability is observed in the decade-long monitoring with the Owens Valley Radio Observatory;][]{Richards2011}\footnote{See  \url{http://www.astro.caltech.edu/ovroblazars/data.php?page=data_return&source=J1148+5924}}, and the optical spectrum and the low polarization ($0.4-0.5\%$), which do not indicate continuum emission other than from stars \citep{Marcha1996}.

\subsection{Origin of the $\gamma$-ray emission}
\begin{figure}
\includegraphics[width=10cm]{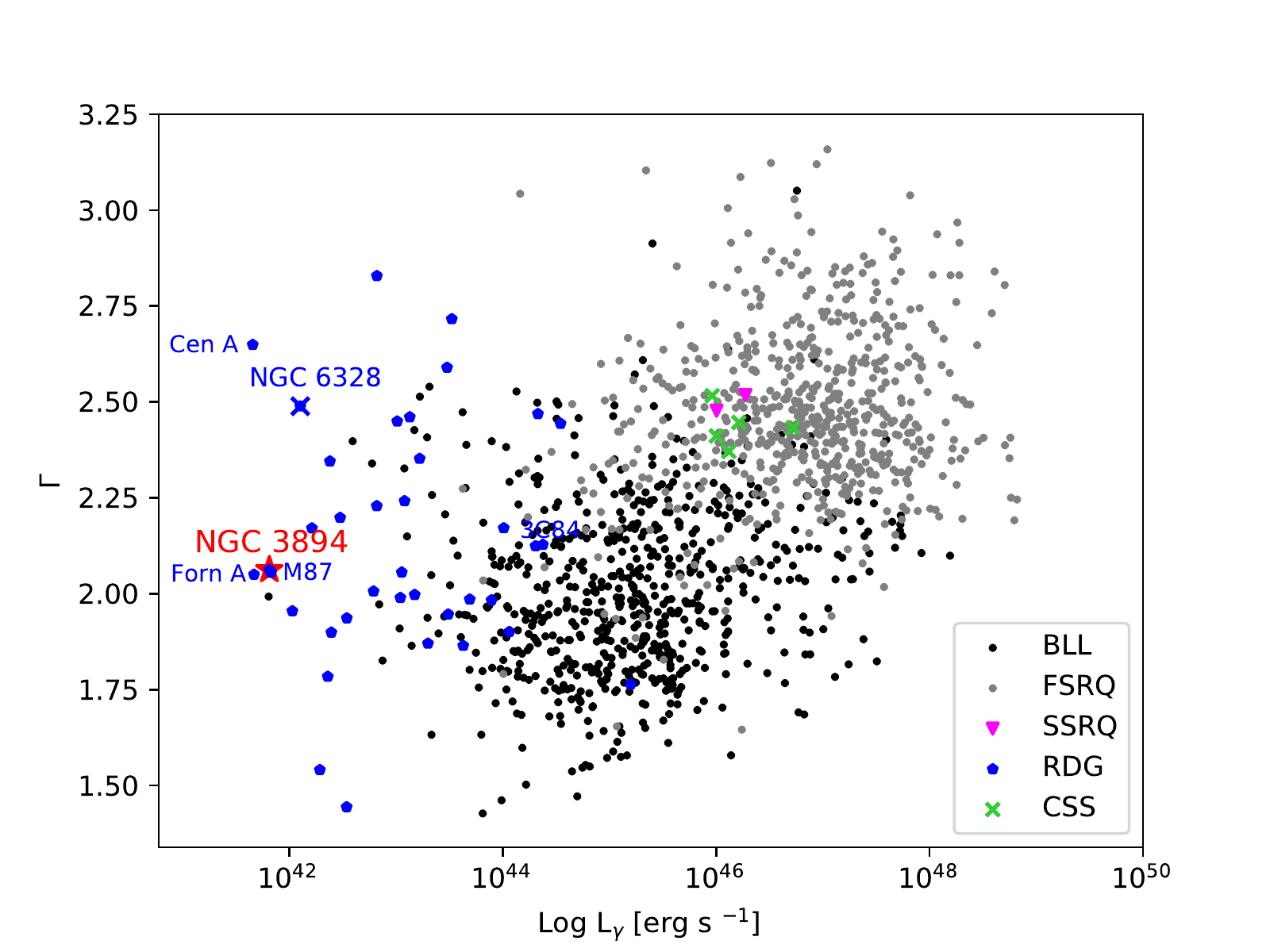}
\caption{\small \label{fig_classification} Diagram of the $\gamma$-ray luminosity vs. spectral index for the extragalactic sources with known redshift contained in the 4LAC. The young radio galaxy NGC\,6328 (also called PKS\,1718-649), which is the first young radio source detected at $\gamma$-rays, is also labeled in the plot.}
\end{figure}

Clues to the origin of the $\gamma$-ray emission in this source can come from a comparison with the other classes of extragalactic $\gamma$-ray emitters.
Figure \ref{fig_classification} shows the diagram of the $\gamma$-ray luminosity versus spectral index for the jetted AGNs with known redshift in the 4LAC.
The high-luminosity side of the plot ($L_{\gamma}\gtrsim 10^{43}$ erg s$^{-1}$) is populated by blazar sources, classified as flat-spectrum radio quasars (FSRQs), BL Lac sources (BLLs), and steep-spectrum radio quasars (SSRQs). Their high luminosities and the frequently observed extreme variability favor an origin of the emission in a compact, highly relativistic region, likely within the aligned inner (subkpc) jet \citep{1995PASP..107..803U,2017A&ARv..25....2P}.
Conversely, radio galaxies (RDGs) are characterized by lower levels of luminosities and variability (with a few notable exceptions), and it is an open question whether this is a consequence of the larger jet inclination angles or of an inherently different site or mechanism for the $\gamma$-ray emission with respect to blazars.

\noindent The $\gamma$-ray luminosity of NGC 3894 is several orders of magnitude lower than that of CSSs, the other intrinsically compact radio objects.
Although the CSSs reported in the 4LAC are much farther away ($z>0.6$), they are all quasars, and their high luminosity suggests that relativistic boosting is likely to play a role in their GeV detection.

NGC 3894 is one of the lowest luminosity ($L_{\gamma}<10^{42}$ erg s$^{-1}$) radio galaxies contained in 4LAC, and it appears to share the same $\gamma$-ray properties as its evolved misaligned counterparts. 
Similar luminosities are also observed for the extended radio galaxies Centaurus A and M87. However, these two are less distant (z$<$0.006), as is the other bona fide young radio source detected by \textit{Fermi}-LAT, NGC\,6328 \citep{2016ApJ...821L..31M}.

The location of the $\gamma$-ray emitting region in radio galaxies is still controversial, with either the jet or the lobes being plausible emitting sites.
At small scales, a clear link between the ejection of bright superluminal knots from the radio core and intense $\gamma$-ray flaring has been observed in the radio galaxies 3C\,111 and 3C\,120 \citep{2012ApJ...751L...3G,2015ApJ...808..162C,2015ApJ...799L..18T}.
On the other hand, at large distances from the nucleus, a firm detection of GeV emission from the extended radio lobes of Cen\,A and Fornax A has been reported in \citet{2010Sci...328..725A} and \citet{2016ApJ...826....1A}.

The most natural object among the misaligned sources for a comparison is the other CSO detected by LAT, NGC\,6328 \citep{2016ApJ...821L..31M}. The two sources have similar (low) $L_{\gamma}$, linear sizes and dynamically estimated ages. The inner structure of NGC\,3894 is as small as $\sim5$ pc and its dynamic age is  $\sim60$ years, whereas the estimated linear size and dynamic age of NGC 6328 are $\sim2$ pc and $\sim100$ yr \citep{2016ApJ...821L..31M}. This seems to suggest that the detection of CSOs is limited to the closest targets and could be connected to the very first evolutionary stages. However, NGC\,3894 has a flatter photon index ($\Gamma = 2.01\pm0.10$) than NGC\,6328 ($\Gamma \sim 2.49\pm 0.18$). Interestingly, this could be related to their different radio morphologies: 
NGC\,6328 shows mini lobes and no clear detection of jets, while NGC\,3894 exhibits a jet (more collimated) structure. It might therefore be speculated that in the former CSO, the $\gamma$-ray emission could come from the compact lobes, as proposed by some models \citep{2008ApJ...680..911S}, whereas it originates from the jet.

Two additional interesting sources for a comparison are the radio galaxies M\,87 (z = 0.0044) and 3C\,84 (also called NGC\,1275, z = 0.0176). 
Both sources have a viewing angle similar to NGC 3894: 15$^{\circ}<\theta<25^{\circ}$ \citep{2016A&A...595A..54M} and $\theta < 18 ^{\circ}$ \citet{2018NatAs...2..472G}, respectively.
The velocities of the jet components are about 0.2c in 3C\,84 \citep{2012ApJ...746..140S} and in the inner part of M\,87 \citep{2016A&A...595A..54M}, similar to what we find for NGC\,3894. 
The three sources have a comparable photon index ($\Gamma$ $\sim$ 2.1). Although it is less distant and more extended ($\sim$100 kpc), M\,87 shows a $\gamma$-ray luminosity of $6-7 \times 10^{41}$ erg s$^{-1}$ , similar to NGC\,3894.
Different is the case of 3C\,84, which has two-sided compact (parsec scale) jets and a distance similar to NGC\,3894, but it presents a $\gamma$-ray luminosity more than 200 times higher ($2 \times 10^{44}$ erg s$^{-1}$).
Unlike NGC\,3894, for which no observations with Cherenkov telescopes have been performed, M\,87 and 3C\,84 are also detected at very high energy (VHE, E $>$ 100\,GeV) \citep{2006Sci...314.1424A,2012A&A...539L...2A}. 
In particular, at VHE, M\,87 displayed strong variability on timescales as short as one day, but no unique signature of the region responsible for the VHE flares has been identified \citep{2012ApJ...746..151A}.
Despite the long-term monitoring of M\,87 during 2012$--$2015 in a low-activity state, the production site of $\gamma$-rays remains unclear \citep{2020arXiv200101643M}. However, the correlation observed between the radio and X-ray activities is a strong indication that most often the emitting region is close to the core in this source.
In the case of 3C\,84, several works on radio, X-ray, and $\gamma$-ray variability suggest that short-term and long-term variability may be produced in different regions of the source. In particular, short-term variability seems related to the injection of fresh particles that are accelerated in a shock in the core region, whereas long-term variability is more likely connected with the jet structure \citep[e.g.,][]{2016arXiv160803652F,2018MNRAS.475..368H}.
Similarly to M\,87 and 3C\,84, long-term radio and X-ray monitoring of NGC\,3894 may provide important information about the origin of the $\gamma$-ray emission.

When we consider the $\gamma$-ray properties of NGC\,3894 together with M\,87, Cen\,A, NGC\,6328, and Fornax\,A, we explor the high-energy emission from a mildly relativistic region ($\Gamma_{bulk}\leq$1.1) and sample the low-power tail of the jet activity.
The case of the two CSOs NGC\,3894 and NGC\,6328 is particularly interesting: beyond being misaligned and low-power radio sources, they represent the first two cases of young ($\lessapprox$100 years) radio sources detected by \textit{Fermi}-LAT.
Comparing these two CSO with two more evolved radio galaxies, Cen\,A and M\,87, which have a similar luminosity $L_\gamma$ and photon index $\Gamma$, we were able to note a possible connection between the photon index and the viewing angle. While Cen\,A \citep{1998AJ....115..960T} and NGC\,6328 have a large viewing angle $\theta >50^{\circ}$ and a steep spectrum $\Gamma>2.5$, NGC\,3894 and M\,87 present a small viewing angle $\theta<25^{\circ}$ and a flat spectrum $\Gamma \sim 2$.

Variable emission observed in the same time interval in radio and $\gamma$-rays can be useful to localize the origin of the high-energy emission in AGN. However, no significant variability has been observed in $\gamma$-rays for NGC\,3894 (see also Fig. \ref{fig_lightcurve}). The lack of variability in $\gamma$-rays is expected for young radio sources with the high-energy emission produced in the lobes.
At radio frequencies, a long-term and dense radio monitoring performed with the Owens Valley Radio Observatory between 2008 and 2019 shows only slow variability of the total flux \citep{2011ApJS..194...29R}. 
For a deep study of the radio emission and morphology variability at milliarcsec scale, we lack observations in the period between 2007 and 2017. 
New VLBA radio observations, together with an extended multiwavelength analysis, including radio, optical, X-ray and $\gamma$-ray data, will be the subject of future work.
In particular, further radio analysis at milliarcsec scale will be essential to continue monitoring the jet expansion and the variability of their components in order to investigate the origin of the $\gamma$-ray emission in this source. Radio observations at arcsec scale may provide important information on the surrounding environments, as well as the connection of the source with large-scale emissions. As previously discussed, X-ray data play a key role in distinguishing between emission mechanisms (accretion or ejection) and in verifying the properties of the environment.
Finally, further studies of the $\gamma$-ray emission will be fundamental for investigating the origin of its high-energy emission.

\subsection*{ACKNOWLEDGMENTS}
The \textit{Fermi}-LAT Collaboration acknowledges generous ongoing support from a number of agencies and institutes that have supported both the development and the operation of the LAT as well as scientific data analysis. These include the National Aeronautics and Space Administration and the Department of Energy in the United States, the Commissariat à l'Energie Atomique and the Centre National de la Recherche Scientifique / Institut National de Physique Nucléaire et de Physique des Particules in France, the Agenzia Spaziale Italiana and the Istituto Nazionale di Fisica Nucleare in Italy, the Ministry of Education, Culture, Sports, Science and Technology (MEXT), High Energy Accelerator Research Organization (KEK) and Japan Aerospace Exploration Agency (JAXA) in Japan, and the K. A. Wallenberg Foundation, the Swedish Research Council and the Swedish National Space Board in Sweden.
Additional support for science analysis during the operations phase is gratefully acknowledged from the Istituto Nazionale di Astrofisica in Italy and the Centre National d'Etudes Spatiales in France. This work performed in part under DOE Contract DE-AC02-76SF00515.
Portions of this research performed at the Naval Research Laboratory (T.J.J., C.C.C.) were supported by NASA
DPR S-15633-Y.
This research made use of the Astrogeo VLBI FITS image database.

\bibliography{main}

\end{document}